\documentclass[aps,pra,twocolumn]{revtex4}   
\newcommand {\be}{\begin{equation}}
\newcommand {\ee}{\end{equation}}
\newcommand {\bey}{\begin{eqnarray}}
\newcommand {\eey}{\end{eqnarray}}
\usepackage{epsfig}
\usepackage{amsfonts}
\usepackage{mathbbol}

\begin{document}

\title{A condition for any realistic theory of quantum systems}

\author{A. Montina}
\affiliation{Dipartimento di Fisica, Universit\`a di Firenze,
Via Sansone 1, 50019 Sesto Fiorentino (FI), Italy}

\date{\today}

\begin{abstract}
In quantum physics, the density operator completely describes the state.
Instead, in classical physics the mean value of every physical quantity is
evaluated by means of a probability distribution. We study the possibility to
describe pure quantum states and events with classical probability 
distributions and conditional probabilities and prove that the distributions can
not be quadratic functions of the quantum state. Some examples are considered.
Finally, we deal with the exponential complexity problem of quantum physics 
and introduce the concept of classical dimension for a quantum system.
\end{abstract}
\maketitle
A peculiar aspect of the standard quantum formalism is interference,
that is, every alternative is not associated with a probability, but with
a complex number. This characteristic is
at the basis of some famous paradoxes as the Schr\"odinger cat~\cite{Sc35}, and
of the "sign problem" encountered in quantum Monte Carlo simulations~\cite{sign_prob},
which gives an exponential growth of the numerical complexity.
Since the birth of quantum mechanics, many physicists investigated the possibility
of its description in terms of classical probabilities.
Indeed, a completely equivalent alternative to the Copenhagen
interpretation is Bohm mechanics~\cite{bohm,bell}, recently extended to quantum
fields with a variable particle number~\cite{durr}. These approaches provide
a realistic description of quantum systems.
Bohm mechanics uses the same mathematical tools of the standard approach, 
as the wave-function, so it differs only in the interpretation and
does not solve for example an important problem as the exponential complexity.
The wave-function is not defined in the three-dimensional physical space,
as a classical field, but in the representation space, thus the number of
variables needed to define the classical state grows exponentially with
the dimension of the physical system. Since quantum mechanics has
a statistical interpretation, there is no a priori
reason to exclude the possibility to describe the quantum states as
statistical ensembles in a smaller space than the Hilbert space.
In some cases, the Wigner function provides such a dimensionality
reduction~\cite{wigner}. It has some properties of a classical probability
distribution in phase-space, but in general it can take negative values.
When it is positive and only particular measurements are performed,
a realistic statistical description in phase-space is possible, as in
the case of 
continuous-variable teleportation experiments involving Gaussian states 
and quadrature measurements~\cite{caves}.

In order to circumvent the sign problem of the Wigner function, some alternative 
non-negative probability distributions were introduced~\cite{pos-prob,wodk},
as the Husimi function. However, also in these cases, a realistic interpretation
is not in general possible. Another example of phase-space distribution is the
P-function, introduced by R. J. Glauber~\cite{quantumoptics}.

A realistic statistical description of quantum mechanics in a
reduced phase-space is very important for its possible 
consequences in Monte Carlo simulations of many-body systems. 
In the cases where it is possible to interpret the Wigner function as 
a probability distribution, as in the truncated Wigner 
approximation, the Monte Carlo approach allows to simulate efficiently the 
dynamics of atoms in degenerate bosonic gases~\cite{trunc_wigner1} and
electrons in semiconductors~\cite{trunc_wigner2}. A general discussion of
the method is also reported in Ref.~\cite{trunc_wigner3}.

In this Letter, we analyze the possibility to have a realistic description
for experiments involving general states and measurements and prove that
a necessary condition has to be fulfilled. Using the non-conflicting
hypotheses of the theorem, we finally introduce the concept of
classical dimension of a quantum system.
In this Letter we use the term "classical" as equivalent to "realistic", 
thus, a classical theory is a theory whose variables have definite values and
for which it is possible to use the classical rules of the probability
theory. In this sense, Bohm mechanics is a classical theory and, consequently, 
every state is classical, i.e., has a realistic description.

We associate 
each quantum state $|\psi\rangle$ with a distribution of probability $\rho(X|\psi)$ in a 
suitable classical space, spanned by a set of variables $X$. This space can be
more general than the phase-space and with higher dimensionality. No a priori 
hypothesis on it is introduced. We want to give physical
meaning to these variables, i.e., we assume that there is an underlying theory
of the quantum system described by them. 
When the system is prepared to a state $|\psi\rangle$, $X$ takes a value with
a probability given by $\rho$.  If a von Neumann measurement is performed, 
the state collapses to a state $|\phi\rangle$ with probability $|\langle\psi|\phi\rangle|^2$.
In terms of the realistic theory, we say that the system with coordinates
$X$ has a conditional probability $P(\phi|X)$ to give the event $\phi$. 
The probability of this event, given the state 
$|\psi\rangle$, is obtained integrating $P(\phi|X)\rho(X|\psi)$ over $X$, as
\be\label{qua_class}
|\langle\phi|\psi\rangle|^2=Tr[\hat P_\phi\hat P_\psi]=\int dX P(\phi|X)\rho(X|\psi),
\ee 
where $\hat P_\psi\equiv|\psi\rangle\langle\psi|$.
The functions $\rho(X|\psi)$ and $P(\phi|X)$ have to satisfy the conditions
\bey\label{pos_cond}
\rho(X|\psi)\ge0 \\
\label{norm_cond}
\int dX\rho(X|\psi)=1,
\eey
\be\label{cond_P}
0\le P(\phi|X)\le1.
\ee
for every $|\psi\rangle$ and $|\phi\rangle$.

A relation as Eq.~(\ref{qua_class}) holds when we evaluate the 
probability distribution of the position of a particle by means of
the Wigner function. The probability to have a particle in the spatial
region $\Omega$ is
\be
\int dq dp P(\Omega|q,p) W(q,p)
\ee
where
\be\label{wign_cond_prob}
P(\Omega|q,p)=\int_{\Omega}\delta(\bar q-q) d\bar q.
\ee
Since $0\le P(\Omega|q,p)\le1$, $P$ it can be interpreted
as a conditional probability. Thus, if the Wigner function
is positive and only position measurements are involved,
a realistic description in phase-space is possible~\cite{caves}.

Every probability distribution introduced so far is quadratic in
the pure quantum state $|\psi\rangle$, as the $P$, the Wigner and the $Q$ 
functions~\cite{quantumoptics}. Indeed, the Wigner function was 
rigorously derived in Ref.~\cite{wigner2} using this property
and other four assumptions.
We will show that quadratic distributions can not be interpreted as 
probability distributions of some realistic theory when general pure 
states and measurements are considered. More precisely,
by assuming that properties~(\ref{qua_class}-\ref{cond_P}) are fulfilled
for {\it every} $\psi$ and $\phi$~\cite{quantum_comp}, we prove that 
the probability distributions associated with pure quantum states are nonlinear 
functions of the density operator. This is the main result of our work.
Note that Bohm mechanics  is a perfectly consistent 
realistic hidden variable theory and the corresponding probability distributions 
are not quadratic in the wave-function, as we will show.

In general, a positive probability distribution linear in the density operator
can be obtained by means of positive operator-valued measurements (POVM)~\cite{povm}. 
We consider a pure state $|\psi\rangle$ and define the associated probability 
distribution with respect to a variable $X$ as 
\be\label{linear_rel}
\rho(X|\psi)\equiv Tr [\hat A(X)\hat P_\psi],
\ee
where $\hat P_\psi\equiv|\psi\rangle\langle\psi|$ is a projector and $\hat A(X)$ is 
a generic Hermitian matrix which depends on $X$. $X$ can be a vector of continuous
and/or discrete variables. $\rho(X|\psi)$ must satisfy the 
properties~(\ref{pos_cond},\ref{norm_cond}).
The first one is fulfilled if $\hat A(X)$ is positive definite, i.e. if
its eigenvalues are non negative. The second one implies that~\cite{povm}
\be\label{normali}
\int dX\hat A(X)=\mathbb{1}.
\ee
After these preliminary remarks, we can prove the theorem. Equations~(\ref{qua_class}-
\ref{cond_P},\ref{linear_rel}) are our starting hypotheses. We will show that they are
conflicting.

From Eqs.~(\ref{qua_class},\ref{linear_rel}), we have
$$
Tr\left\{[\int dX P(\phi|X)\hat A(X)-\hat P_\phi]\hat P_\psi\right\}=0,
$$
for every $|\psi\rangle$ and $|\phi\rangle$. Thus,
\be\label{proje}
\hat P_\phi=\int dX P(\phi|X)\hat A(X).
\ee
Since $\hat A(X)$ is positive definite and $P(\phi|X)\ge0$, it is evident that 
\be\label{XvsP}
P(\phi|X)\ne0\Rightarrow  \hat A(X)\propto\hat P_\phi.
\ee
Thus, if $P(\phi|X)\ne0$ and $P(\phi'|X)\ne0$,
then $\phi=\phi'$. We can define the following one-valued function
\be
\chi: X\rightarrow \chi(X) \text{ such that } P(\chi(X)|X)\ne0.
\ee
The function $\chi(X)$ spans the entire Hilbert space, but it is not
necessarily invertible. It is possible to introduce an auxiliary
set $Y(X)$ of variables in order to have the bijective mapping 
$X\leftrightarrow(\chi,Y)$. Condition~(\ref{XvsP}) implies that
\be\label{zero_P}
P(\phi|\chi,Y)=0 \text{  if  }\phi\ne\chi 
\ee
\be\label{A_proje}
\hat A(\chi,Y)\equiv\alpha(\chi,Y)\hat P_\chi,
\ee
where $\alpha(\chi,Y)\ge0$.
Equations~(\ref{proje},\ref{zero_P},\ref{A_proje}) give
\be\label{tot_pro}
\int {\cal D\chi} dY P(\phi|\chi,Y)\alpha(\phi,Y)=1,
\ee
for every $\phi$. Since $P(\phi|\chi,Y)$ is different from zero in a set 
with zero measure [Eq.~(\ref{zero_P})] and it is smaller or equal to 1,
$\int dY\alpha(\phi,Y)$ must be infinite for every $\phi$, but
this is impossible, because of the normalization 
condition~(\ref{normali}). \ $\square$

Since the properties~(\ref{pos_cond}-\ref{cond_P}) are the minimal necessary
conditions for a realistic theory, we have to discard the linear assumption,
i.e., Eq.~(\ref{linear_rel}). This is sufficient to remove the contradiction,
as shown below, when we will explicitly write the probability density and 
conditional probability for the Bohm theory.

Now, we consider some examples to illustrate our result.
The $P$, Wigner and $Q$ functions can be defined for systems described by
boson creation and annihilation operators, $\hat a^\dagger$ and $\hat a$,
respectively~\cite{quantumoptics}.
It well-known that the $P$ and Wigner function cannot be probability
distributions. The first one is highly singular for particular states,
as squeezed states or superposition of coherent states. The second
one is well-defined for every state, but can assume negative values,
i.e. equation~(\ref{pos_cond}) is not satisfied. The $Q$-function requires
a more detailed discussion. It is smooth, positive and normalized, thus
our theorem says that equation~(\ref{cond_P}) cannot be satisfied.
Consider a one-mode system and denote the $Q$-function corresponding
to a state $|\psi\rangle$ by $Q(\alpha|\psi)$, where $\alpha$ is
a complex number. The expectation value of an observable $M(\hat a^\dagger,\hat a)$
in anti-normal form is given by the average of the function $M(\alpha^*,\alpha)$
with respect to $Q(\alpha|\psi)$, The expectation value of the number of
particles is 
$\langle \hat a^\dagger\hat a\rangle=\langle \hat a\hat a^\dagger\rangle-1 
=\int d^2\alpha(|\alpha|^2-1)Q(\alpha|\psi)$,
thus we have that $\sum_{n=0}^\infty n P(n|\alpha)=|\alpha|^2-1$,
where $P(n|\alpha)$ is the conditional probability of measuring n particles for the
phase space state $\alpha$. This equation implies that the conditional probability
has to be negative for some values of $n$ and $|\alpha|^2<1$, in agreement with
our theorem. 

Now, we consider the case of a two dimensional Hilbert space. It is possible
to construct a probability distribution $\rho(n|\psi)$ with the 
following $3$ matrices~\cite{ahnert}:
$\hat A_1=\frac{1}{3}(\mathbb{1}+\hat\sigma_3)$, $\hat A_2=\frac{1}{3}\mathbb{1}
-\frac{1}{6}\hat\sigma_3+\frac{\sqrt{3}}{6}\hat\sigma_1$ and
$\hat A_3=\frac{1}{3}\mathbb{1}
-\frac{1}{6}\hat\sigma_3-\frac{\sqrt{3}}{6}\hat\sigma_1$, where $\hat\sigma_i$ are
the Pauli matrices.
The distribution has three values, is positive and normalized. Also in this case,
the conditional probability $P(\phi|n)$ cannot satisfy Eq.~(\ref{cond_P}).
For example, when $\phi=(1,0)$, Equation~(\ref{qua_class}) is fulfilled if and only
if $P(\phi|1)=3/2$ and $P(\phi|2)=P(\phi|3)=0$.

We can define another probability distribution as
$\rho(\theta,\phi|\psi)=\frac{1}{2\pi}Tr[|\theta,\phi\rangle\langle\theta,\phi|\hat\rho]$,
where $|\theta,\phi\rangle=\cos(\theta/2)|0\rangle+e^{i\phi}\sin(\theta/2)|1\rangle$
and $|0\rangle$, $|1\rangle$ are two orthonormal vectors. $\theta$ and $\phi$
are the polar coordinate of a point of the Bloch sphere. In the integrals with respect
to these coordinates, we use the measure $\sin\theta d\theta d\phi$. The distribution
is positive and normalized. For $|\psi\rangle=|0\rangle$, we have 
$\rho(\theta,\phi|0)=[\cos(\theta/2)]^2/2\pi$. The probability of obtaining
the state $|0\rangle$ is equal to $1$, thus
\be
\int_0^{2\pi} d\phi\int_0^\pi \sin\theta P(0|\theta,\phi)\rho(\theta,\phi|0)=1
\ee
We assume ab absurdo that $P(0|\theta,\phi)$ is positive, then the equality is satisfied 
only if $P(0|\theta,\phi)=1$, for $\theta\ne\pi$. But this implies that the probability 
to obtain $|0\rangle$ when the system is in the state $|1\rangle$ is different
from zero, since the corresponding probability distribution 
$\rho(\theta,\phi|1)=[\sin(\theta/2)]^2/2\pi$
is zero only for $\theta=0$. This is absurd because
$|0\rangle$ and $|1\rangle$ are orthogonal. 
All the previous examples show that at least one property required by the theorem is
not satisfied. 

Discarding the linear hypothesis, we investigate how to construct a 
probability distribution associated with every density operator which satisfies the
conditions~(\ref{qua_class}-\ref{cond_P}). The Bohm mechanics provides a simple 
example of true probability. For sake of simplicity, here we consider only
the case of a fixed number of particles~\cite{bohm,bell}. Recently,
the theory has been extended to quantum fields and accounts for creation
and annihilation of particle~\cite{durr}.
The dynamical variables are a multi-particle wave-function $\chi$
and the coordinates $\vec x$ in the configuration space. If the quantum system
is in a pure state $\psi$, the variable $\chi$ is merely equal to $\psi$
and the distribution of the coordinates $\vec x$ is $|\psi(\vec x)|^2$, thus
\be
\rho(\vec x,\chi|\psi)\equiv |\psi(\vec x)|^2\delta(\chi-\psi),
\ee
where $\psi$ is the state in the Hilbert space of the quantum system.
This function is obviously non-quadratic in $\psi$, because of the
Dirac delta. 
In Bohm mechanics, the position measurement gives the variables $\vec x$
as result, thus, the conditional probability of finding the system in a volume 
$\Omega$ in the representation space is
\be\label{cond_bohm}
P(\Omega|\vec x,\chi)=\int_{\Omega}d\vec x_0\delta(\vec x_0-\vec x).
\ee
Other measurements, for example of momentum, can be performed by
a suitable unitary evolution and a subsequent measurement of position.
During the evolution, the coordinates $\vec x$ go to a new value 
which is function of $x$ and $\psi$. This enables to obtain a positive
conditional probability for every measurement from Eq.~(\ref{cond_bohm}).

Bohm mechanics is a free-dispersion theory, i.e., the hidden
variables fix exactly the results of measurements. It corresponds
to have conditional probabilities equal to $0$ or $1$.
This characteristic is not necessarily required by a realistic theory.
Indeed, the minimal requirements are Eqs.~(\ref{qua_class}-\ref{cond_P}).
Discarding the free-dispersion hypothesis, the simplest probability distribution 
for a pure state $\psi$ is 
\be\label{simplest}
\rho(\chi|\psi)=\delta(\chi-\psi),
\ee
where $\chi$ is the variable of the classical system.
The corresponding conditional probability for the event $\phi$ is 
$P(\phi|\chi)=|\langle\phi|\chi\rangle|^2$.

Although these two examples sound trivial, they show that properties
(\ref{qua_class}-\ref{cond_P}) are not
conflicting. At this point, we raise a question regarding the exponential 
complexity of quantum mechanics.
In the case of Eq.~(\ref{simplest}), the dimension of the phase space
spanned by the variables $\chi$ is obviously the dimension of
the Hilbert space of $\psi$. It is well-known that the dimension of the Hilbert
space grows exponentially with the physical dimension of the system. For
example, it is $2^N$ for $N$ spins $1/2$.
Thus, also the number of variables $\chi$ of the classical theory has an 
exponential growth. Every known hidden variable theory has this
feature. However, if we discard some required property, as 
positivity, the dimension of the $X$ space can be considerably reduced.
For example, the Hilbert space dimension of one boson mode is infinite,
since the number of particles goes from zero to infinity, but
the corresponding Wigner functions have only two variables, although
we cannot regard these variables as describing a classical
system, because of the negativity of the Wigner function.
In general, for a Hilbert space with dimension $M$, 
it is possible to find quasi-probability distributions where $X$ can assume
only $M\times M$ values, that is, if the Hilbert space dimension 
is finite, we can have a space of $X$ with dimension equal to zero.
So, on one hand we have an example of true 
probability distribution on a space with the same dimension of
the Hilbert space, on the other one, we can have quasi-probability
distributions on a space with a considerably lower dimension
with respect to the Hilbert space. The question we raise is the
following: if the dimension of the Hilbert space is $D_{quant}$,
which is the lowest dimension $D_{class}$ for $X$ in 
order to fulfill the conditions~(\ref{qua_class}-\ref{cond_P})?
Equation~(\ref{simplest}) says that $D_{class}\le D_{quant}$.
The introduction of a probability distribution $\rho(X|\psi)$ and
a conditional probability $P(\phi|X)$ for each event
allows to place the question of the exponential complexity nature
of quantum mechanics onto a clear and well-defined ground and
the proof that $D_{class}=D_{quant}$ or that $D_{class}\ll D_{quant}$ 
would cast new light upon this question.
The evaluation of $D_{class}$ is not a trivial problem and there is
no evident reason for assuming it equal to $D_{quant}$. We have demonstrated
that the solution of the problem requires to discard probability distributions
quadratic in the wave-function.
Note that we have not dealt with dynamical considerations, but only with 
the problem of a classical representation of quantum states. 

In conclusion, we have studied the possibility of a classical description
of quantum states and events by means of probability distributions and
conditional probabilities, that must satisfy the 
properties~(\ref{qua_class}-\ref{cond_P}). 
We have demonstrated
that the probability distribution for a pure state has to be a nonlinear 
function of the density operator. We have illustrated the proof
with some examples, as the $P$, Wigner and $Q$ functions, and two
cases of POVM for a two-state system. We have shown that
for Bohm mechanics the distributions of probability are not quadratic
functions of the pure quantum state, as required by our theorem to any
realistic theory. Finally, the concept of classical
dimension $D_{class}$ of a quantum system has been introduced.
Every known hidden variable theory has not a phase-space dimension
smaller than the Hilbert space dimension $D_{quant}$. 
We conclude with a non-trivial question.
Does a classical theory exist whose phase space dimension is much 
smaller than $D_{quant}$? that is, is $D_{class}\ll D_{quant}$?
Bohm mechanics could be considered as a pure interpretation
curiosity, since it uses essentially the same tools of quantum mechanics,
whereas a theory with much lower dimensionality would have
important implications.
We have put this problem onto a well-defined ground and
its solution could clarify the nature of the exponential complexity
of quantum mechanics.

I thank F. T. Arecchi for helpful discussions. This work was supported
by Cassa di Risparmio di Firenze under the project "dinamiche cerebrali caotiche".

\end{document}